%% file: main.tex
\documentclass[]{cas-sc}
\usepackage[backend=bibtex,sorting=none,maxcitenames=1000,maxbibnames=99,style=numeric-comp]{biblatex}

\usepackage{cleveref}
\usepackage{svg}
\usepackage{graphicx}
\usepackage{amsmath,amsthm, amssymb, float, bm, subcaption}
\usepackage{placeins}
\usepackage{longtable,tabularx,booktabs}
\usepackage{soul}
\usepackage{tcolorbox}
\usepackage{placeins,fancyhdr,tikz,todonotes}
\definecolor{C0}{HTML}{1F77B4}
\definecolor{C1}{HTML}{FF7F0E}
\definecolor{C2}{HTML}{2ca02c}
\definecolor{C3}{HTML}{d62728}
\definecolor{C4}{HTML}{9467bd}
\definecolor{C5}{HTML}{8c564b}

\makeatletter                                                                   
\newlength{\bibsep}{\@listi \global\bibsep\itemsep \global\advance\bibsep by\parsep} 
\makeatother

\shorttitle{}    
\shortauthors{}  
\title [mode = title]{The Alamo multiphysics solver for phase field simulations with strong-form mechanics and block structured adaptive mesh refinement}  
\author[isu]{Brandon Runnels}[orcid=0000-0003-3043-5227]
\cormark[1]
\cortext[1]{Corresponding author}
\author[lanl]{Vinamra Agrawal}[orcid=0000-0002-1698-1371]
\author[asu]{Maycon Meier}[orcid=0000-0001-9914-8811]
\affiliation[isu]{organization={Department of Aerospace Engineering, Iowa State University},
city={Ames},
state={IA},
country={USA}}
\affiliation[lanl]{organization={Materials and Physical Data Group, Los Alamos National Laboratory},
city={Los Alamos},
state={NM},
country={USA}}
\affiliation[asu]{organization={Department of Materials Science and Engineering, Arizona State University},
city={Temple},
state={AZ},
country={USA}}

\begin{document}
\begin{abstract}
  Alamo is a high-performance scientific code that uses block-structured
  adaptive mesh refinement to solve such problems as: the ignition and
  burn of solid rocket propellant, plasticity, damage and fracture in
  materials undergoing loading, and the interaction of compressible flow
  with eroding solid materials. Alamo is powered by AMReX, and provides a
  set of unique methods, models, and algorithms that enable it to solve
  solid-mechanics problems (coupled to other physical behavior such as
  fluid flow or thermal diffusion) using the power of block-structured
  adaptive mesh refinement.
\end{abstract}

\maketitle

\input{paper.tex}
\printbibliography

\end{document}

%% file: paper.tex
\hypertarget{statement-of-need}{%
\section{Statement of need}\label{statement-of-need}}

The phase field (PF) method is a powerful theoretical framework that
enables the systematic description of complex physical systems
\autocite{qin2010phase,steinbach2009phase,burger2006phase}. PF methods
have been successfully used to describe phenomena such as
solidification, microstructure evolution, fracture, damage,
dislocations, and many more. Beyond materials science, PF methods have
also enjoyed great success in other applications ranging from
deflagration of solid rocket propellant to topology optimization.

The success of the PF method is derived from its implicit, diffuse
representation of boundaries and surfaces, which avoids the need for
cumbersome interface tracking. However, the PF method also can incur
great computational expense, due to the need for high grid resolution
across the diffuse boundary. In order for the PF method to be feasible,
strategic algorithms are necessary in order to provide sufficient
boundary resolution without wasting grid points on uninteresting
regions. Such algorithms fall typically into two main categories. (1)
Spectral methods solve the phase field equations in the frequency
domain, e.g. \autocite{kochmann2015phase}, and (2) Real-space methods
employing adaptive mesh refinement (AMR). Spectral methods offer a
number of performance advantages, especially when coupling to global
mechanical solvers. However, they can be limited in their ability to
resolve fine-scale features, and can be very cumbersome to use when
implementing novel types of models. On the other hand, real-space
methods with AMR are often able to attain very good performance, can be
easily suited to the domain of interest, and provide an attractive
platform for prototyping new physical models.

A number of open-source real-space PF codes exist and have enjoyed
significant popularity. Some of the most widely known codes with PF
implementations are Moose, \autocite{giudicelli2024moose}, Fenics
\autocite{baratta2023dolfinx}, and Prisms-PF
\autocite{dewitt2020prisms}, which employ octree style AMR. Though
effective, octree-AMR can result on complex and expensive mesh
management. It is can also be challenging to achieve optimal load
balancing, due to the high degree of unpredictable connectivity within
the octree mesh.

Block-structured AMR (BSAMR) is an alternative AMR strategy. BSAMR
divides the domains into distinct levels, with each level usually
consisting of a collection of Cartesian grid regions (patches), that
effectively evolve independently. Communication between patches and
levels is then handled through ghost cells, interpolation, and
restriction. This data structure is extremely efficient and scalable,
while also being highly amenable to efficient code prototyping.
Importantly, BSAMR also acts as a seamless extension to geometric
multigrid, making naturally efficient at performing global mechanical
equilibrium solves. The AMReX framework \autocite{zhang2019amrex}
provides a powerful platform for development of BSAMR codes. However,
the use of AMReX (and BSAMR in general) has been limited in PF and solid
mechanics, due to the inherent challenges of solving the mechanical
equilibrium equations on a patch-based mesh.

The Alamo multiphysics solver leverages the power of BSAMR for
phase-field problems. Alamo provides a unique, strong-form
finite-deformation, matrix-free mechanics solver, enabling the efficient
solution of the solid mechanics calculations. It also provides a set of
numerical integration routines, myriad material models, and numerous
examples covering a broad cross-section of PF modeling interests.

\hypertarget{methods}{%
\section{Methods}\label{methods}}

\hypertarget{mechanical-solver}{%
\subsection{Mechanical solver}\label{mechanical-solver}}

The Alamo mechanical solver extends the multi-level multi-grid (MLMG)
solver to address the problem of quasi-static mechanical equilibrium,
that is, \begin{equation} \label{eq:mech_equil}
\operatorname{Div}\big( \mathrm{DW}(\mathbf{F})\big) + \mathbf{B} = \mathbf{0},
\end{equation} where \(\mathbf{F}\) is the deformation gradient,
\(\mathbf{B}\) is a body force, and \(\mathrm{W}\) is an arbitrary
Helmholtz free energy with derivatives
\(\mathbf{P}=\mathrm{DW}(\mathbf{F}) = d\mathrm{W}/d\mathbf{F}\) the
Piola-Kirchhoff stress tensor, and
\(\mathbb{C}=\mathrm{DDW}(\mathbf{F})=d^2\mathrm{W}/d\mathbf{F}/d\mathbf{F}\)
the tangent modulus. Usually, the solution of \autoref{eq:mech_equil} is
achieved using the finite element method (FEM). However, the main
advantages of FEM (conformal meshing, ability to handle material
discontinuities) are not relevant in the BSAMR framework. Moreover, FEM
often presents difficulties in achieving consistent shape functions
between levels, causing problems in achieving consistent derivatives at
the coarse/fine boundary without a global matrix.

The Alamo solver, on the other hand, is developed to be native to the
BSAMR framework, and takes full advantage of the integration with
geometric multigrid. It is matrix-free, which is necessary in order to
avoid additional communication overhead. It is strong-form, using finite
differences instead of shape functions to calculate derivatives,
ensuring consistency between levels and compatibility with
restriction/prolongation operations. It handles coarse-fine boundaries
using a novel ``reflux-free'' method, which avoids the special treatment
of boundaries by including an extra layer of smoothed ghost nodes.
Details on these aspects of the solver are available in
\autocite{runnels2021massively}.

The Alamo mechanical solver is versatile, allowing any type of
mechanical model to be used though the abstract solid model interface
(per the norm for most FEM codes). Alamo uses templated AMReX BaseFab
structures to encapsulate model parameters, which enables users to
implement sophisticated models without any knowledge of the external
Alamo/AMReX infrastructure. An optional interface is provided that
allows users to communicate model information to the Alamo I/O routines,
in order to include internal solid mechanics variables (such as
accumulated plastic slip) with the field output. Because mechanical
models are instantiated as field variables, they require arithmetic
operations in order to be interpolated/averaged as the mesh is adapted.
Alamo implements a ``model vector space'' framework that provides an
intuitive and efficient framework for automatic creation of arithmetic
operations. This allows model developers to implement all necessary
arithmetic operators with only three additional lines of code, ensuring
systematic compliance of the model and increased readability of the
code. The flexibility of this framework is exemplified through Alamo's
library of solid models, which include implementations of solids ranging
from linear elastic isotropic materials to finite deformation strain
gradient crystal plasticity.

Because the mechanical solver is coupled to problems defined with
diffuse boundaries, often involving ``void'' regions in which there is
no mechanical strength, additional steps are necessary to avoid
convergence issues. Alamo contains methods for accounting for diffuse
boundary conditions, and uses a joint cell/node based interpolation
scheme to ensure good convergence even when the operator is
near-singular. Details on the near-singular solver capability, as well
as the model vector space implementation, have been documented in
\autocite{agrawal2023robust}.

\hypertarget{multiple-inheritance-polymorphic-integrators}{%
\subsection{Multiple inheritance polymorphic
integrators}\label{multiple-inheritance-polymorphic-integrators}}

In Alamo, each type of physical behavior is encapsulated by an
``Integrator'' class. Each integrator is responsible for solving a
certain set of equations: the HeatConduction integrator solves the
thermal diffusion equation; the AllenCahn integrator solves the
Allen-Cahn equations; the Mechanics integrator solves the equations of
mechanical equilibrium; etc. All integrators inherit from the base
integrator, which interfaces with AMReX and manages the
creation/deletion/evolution of field variables. This partitions the code
so that each integrator is responsible only for its own physics, without
excessive bookkeeping infrastructure.

Most PF problems of interest feature the non-trivial interaction of
multiple disparate physical behaviors. For example, microstructure
evolution in metals is strongly coupled to mechanical loading. Or, some
phase field models may require a fully resolved flow-field. Such
couplings can rapidly increase the complexity of the code base.

The Alamo solution is \textbf{multiple-inheritance polymorphic
integrators} (MIPI). The MIPI schema encapsulates each physical system
into its own integrator. Single-application integrators inherit directly
from the base integrator, whereas multi-physics application link other
integrators together using multiple inheritance.

\begin{figure}
\centering
\includegraphics[width=0.5\textwidth]{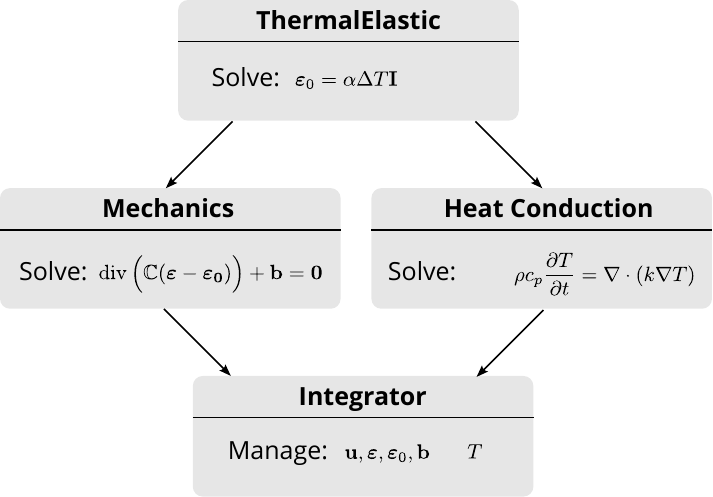}
\caption{The MIPI schema for solving
thermoelasticity.\label{fig:example}}
\end{figure}

The MIPI method is exemplified through the case of coupled thermal
evolution and mechanical equilibrium (\autoref{fig:example}). In this
example, the Mechanics integrator is only responsible for solving
mechanical equilibrium and can be run independently. Similarly, the Heat
Conduction integrator is concerned only with thermal evolution; all
physical aspects related to heat conduction are encapsulated here. Both
Mechanics and Heat Conduction inherit from the integrator base class,
which tracks and updates all field variables on a shared grid. To link
these together, the ThermalElastic integrator inherits jointly from
Mechanics and HeatConduction. Its only function is to link the Mechanics
strain field to the Heat Conduction temperature field; everything else
is managed by the respective integrator. This allows for integrators to
be combined in arbitrary ways, with minimal (or possible no)
instrumentation required within the linked integrator itself.

Numerous alamo integrators have been developed and used for scientific
applications. A brief summary is included here:

\begin{itemize}
\item
  Microstructure evolution is simulated using the multi-phase field
  method \autocite{eren2022comparison} combined with the strong-form
  mechanics solver to simulate grain boundary anisotropy
  \autocite{ribot2019new}, phase field disconnections
  \autocite{runnels2020phase,gokuli2021multiphase}, and twin growth in
  magnesium \autocite{hu2024atomistic}.
\item
  The phase field fracture mechanics model couples crack evolution and
  mechanics to capture crack propagation
  \autocite{agrawal2021block,agrawal2023robust}.
\item
  Deflagration of solid rocket propellant is captured using a phase
  field method \autocite{kanagarajan2022diffuse}, coupled to heat
  transfer \autocite{meier2024diffuse} and the hyperelastic solver
  \autocite{meier2024finite} to accurately predict ignition,
  deflagration, and mechanical response.
\item
  Recent work demonstrates the ability of Alamo to simulate a
  hydrodynamic compressible flow through a domain governed by phase
  field equations including the Allen Cahn equation (for porous media)
  and the dendrite growth equation \autocite{boyd2025diffuse}.
\end{itemize}

\hypertarget{infrastructure}{%
\section{Infrastructure}\label{infrastructure}}

Alamo is intended to bridge the gap between \emph{prototype/research
codes} (often written with \emph{ad-hoc} structure in a non-scalable
language with minimal documentation) and \emph{production codes} (which
are well-organized and documented, but have a steep learning curve and
restrictive contribution requirements). In other words, it is designed
to be easy for an inexperienced researcher to learn, contribute to, and
run without excessive time spent learning the infrastructure. Automation
is therefore the key to ensure that Alamo retains production-level
quality without requiring excessive effort on the part of the
developers.

\hypertarget{parameter-parsing-and-automatic-documentation}{%
\subsection{Parameter parsing and automatic
documentation}\label{parameter-parsing-and-automatic-documentation}}

Alamo employs a recursive object-oriented parameter parsing system. Each
class contains a standard \texttt{Parse} function to read its member
variables, and is located as close as possible to the location in the
code as where the variables are used. Parser commands self-document the
nature of each variable: for instance, \texttt{query\_default} provides
a default value, \texttt{query\_required} triggers an error if the
variable is not provided, \texttt{query\_validate} provides a list of
acceptable values, and so on. This is effective at eliminating
uninitialized variables, and localizes the variable's usage with its
parsing statement. All parameters are read in hierarchically,
systematically using prefixes to avoid naming conflicts (especially in
the case of MIPI integrators).

In addition, Alamo contains a set of python-based code scrapers to scan
the source code for all input parameters. The inputs are cataloged,
along with their source code location and comment string, and formatted
into the automatically-generated documentation. This allows users to
browse the inputs and link directly to their usage in the code (a common
difficulty when encountering a new code), or to search an input index to
determine how the inputs are used. It is also used to automatically
generate input file builders that are guaranteed to be accurate and
consistent with the current source code.

Continuous integration is used to require that all inputs be adequately
documented. Then, the documentation is automatically generated and
posted online with every addition to the main development branches. This
ensures that all documentation is kept current with the source code,
without requiring anything of the developer beyond a single comment
string for each input.

\hypertarget{automatic-regression-testing-system}{%
\subsection{Automatic regression testing
system}\label{automatic-regression-testing-system}}

Regression tests are essential to ensure reliability of a continuously
developed code. Alamo has a self-contained regression test system
designed to run with a single line of code added to an input file stored
in the repository. It also contains a suite of python helper functions
that use the YT library to automatically extract data for comparison and
to determine whether runs have completed accurately
\autocite{turk2010yt}. The regression test system is automated using the
Github Actions CI system, and different variants of tests are executed
automatically in different stages of code development. For example,
full-scale tests are run on a self-hosted runner in each of the main
development branches; lightweight suite of tests are run upon every
feature branch commit. Selections of the tests are used to check code
coverage using gcov, and memory safety using ASan and MSan.

Regression tests are beneficial also for providing starting points for
new users and developers of the code. Alamo's automatic documentation
system integrates with the automatic test system to generate
comprehensive documentation for each test, including figures and run
instructions. Since it is implemented with CI, the online test
documentation is guaranteed to be up-to-date.

\hypertarget{guaranteed-reproducibility}{%
\subsection{Guaranteed
reproducibility}\label{guaranteed-reproducibility}}

A hallmark feature of Alamo is its guaranteed reproducibility system.
Alamo integrates with the AMReX ParmParse system to track all input
parameters. On every execution, Alamo creates a metadata file in the
output directory that stores all input parameters, as well as platform
information, git commit ID, amrex git commit ID, etc.

Because Alamo is a development code, sometimes last-minute modifications
are made to the source that are not committed prior to the code's
execution. To resolve this, as part of the build process, Alamo always
runs a git diff on its own source code, and stores the results. When the
code is subsequently executed, the git diff is stored with the output,
ensuring that future users can always revert the code back to the exact
same state at a later time. This guarantees that every Alamo result,
regardless of the state of the code, is reproducible.

\hypertarget{acknowledgments}{%
\section{Acknowledgments}\label{acknowledgments}}

The authors acknowledge the many funding sources that have supported the
development of Alamo. This includes: Support from Lawrence Berkeley
National Laboratory, subcontracts \#7473053, \#7645776; National Science
Foundation, grants \#OAC-2017971, \#MOMS-2142164, \#MOMS-2341922; and
the Office of Naval Research, grants \#N00014-21-1-2113,
\#N00014-25-1-2029.